# Chip-scale monolithic optoelectronic voltage boost conversion

Parthiban Santhanam*, Daniel Cui, Jae S. Hwang, David Abraham, Isabella He, Enzo Watanabe and Aaswath P. Raman**

Department of Materials Science and Engineering, University of California, Los Angeles, Los Angeles, CA USA 90095

California NanoSystems Institute, University of California, Los Angeles, Los Angeles, CA USA 90095

**Abstract**

Voltage conversion is a fundamental electronic process critical to engineered systems across a wide spectrum of applications and spanning many orders of magnitude in scale. Conventional approaches like transformers and charge pumps perform well in specific contexts but face fundamental limitations to miniaturization, electromagnetic interference, and voltage range. Here we present a chip-scale, fully integrated monolithic, non-switching optoelectronic voltage conversion platform capable of high gain, bootstrap-free boost-mode operation across several orders of magnitude in power density and voltage scale. Using the bidirectional coupling between LEDs and PV cells with identical active layer materials, our chip-scale, single-die strategy eliminates Stokes losses while improving key parameters like physical footprint, series resistance, and photon leakage by orders of magnitude over implementations using multiple packaged, discrete components. Moreover, by exploiting the large étendue of NIR-transparent semi-insulating InP substrates and the atomically smooth, void-free interface of lattice-matched epitaxial growth, simulations indicate that our InGaAsP architecture's photon transport simultaneously provides a >60× increase in current density and >50× reduction in non-radiative recombination losses compared with a multiple-die solution while simultaneously reducing fabrication complexity and improving mechanical robustness. We experimentally demonstrate a boost gain of 3.8× in an 8x8 mm$^2$ InGaAs-on-InP chip while validating key aspects of the voltage conversion platform.

\*   parthi@g.ucla.edu
\*\* aaswath@ucla.edu

**Introduction**

Voltage conversion is a basic functionality of power electronic circuits deployed across a broad range of critical technologies ranging from grid-scale high-voltage DC (HVDC) electricity transmission and electric vehicles (EVs) to compact power supplies for embedded sensor nodes and energy-efficient multi-core chipsets. When the capacity to increase the voltage at which power flows (i.e., voltage boost) is required, a range of approaches may be employed to achieve this. At the largest scales, magnetic induction transformer-based voltage boost converters have been optimized to minimize current-driven Joule heating losses in transmission lines[1,2]. At the chip-scale complementary metal oxide semiconductor (CMOS) charge pumps are used to mate fluctuating, intermittent and often low power sources to digital and biomedical systems in need of stable volt-scale power[3,4]. However, these approaches face certain fundamental limitations that limit the scope and range of where voltage conversion is accessible.

Most voltage conversion technologies, including transformers and charge pumps, are fundamentally dependent on oscillating or switching phenomena that are not strictly necessary for direct-current-to-direct-current (DC-to-DC) voltage conversion. Switching electronic components produce electro-magnetic interference (EMI) and lose power to various parasitic processes away from base-band[5]. For DC-DC voltage converters, the list of drawbacks expands to include the introduction of voltage ripples that can contaminate downstream circuits[6].

Magnetic transformers, while common at the largest scales, are difficult to miniaturize because they rely on 3-dimensional coil structures and sizeable volumes of magnetic materials that are difficult to integrate at chip-scale[7–9]. Although innovations like additive manufacturing are being studied to address associated issues like manufacturing costs[10,11], transformers in general remain ill-suited to small-scale systems. Charge pumps on the other hand are small, lightweight, inexpensive CMOS-manufacturable circuits[12–14] that convert voltage by repeatedly switching the topology of an on-chip capacitor network. However, boost converters of this type face challenges at low voltages, because performing these changes requires enough starting voltage to drive the CMOS circuit elements that perform the switching.

If the input voltage is too low to switch a Silicon transistor, higher voltages found elsewhere on the chip during steady-state operation may not initially be available and an explicit start-up or bootstrap process is required. Bootstrapping a DC-DC boost converter, whether by conventional battery- or capacitor-based approaches[12,14] or moving parts[13,15,16] reduces robustness, deployment flexibility, and reliability. Bootstrapping also increases complexity, size, and cost while fundamentally introducing a start-up delay that precludes their use in time-sensitive applications. Beyond the input voltage hurdles, charge pumps also face on-chip capacitance requirements that impose trade-offs between miniaturization, voltage ripple, and output power.

An alternative approach to voltage boost conversion that holds the potential to overcome these limitations is to leverage optoelectronics by illuminating photovoltaic cells (PVs) with light-emitting diodes (LEDs). Recent work demonstrated this approach at the board scale[17] and similar concepts are employed by commercial multi-chip optocouplers that boost voltage but which prioritize optical signal isolation over power conversion performance[18–22]. While elucidating, these past works face a fundamental performance limitation due to Stokes losses inherent to their design. Additionally, they have been limited to bulky electronic components making them unsuitable for a broad range of chip-scale devices that need voltage converters. Collectively, a monolithic, chip-scale approach to voltage boost conversion that circumvents the limitations of the switching architectures would constitute a fundamentally new power electronic capability but is unavailable today.

In this Article, we present a photonic integrated circuit (PIC) platform for chip-scale DC-DC voltage conversion capable of non-switching, high gain boost-mode operation in a manner that overcomes many of the fundamental limitations of existing converters. Our approach relies on monolithically harnessing an optoelectronic approach wherein a low input voltage drives one large-area LED which simultaneously illuminates many smaller, series-connected PV cells to generate a high output voltage. We design, fabricate, and characterize a 64 mm$^2$ InGaAs-on-InP chip with >100 devices that demonstrates the viability of a new monolithic circuit architecture. Our converter chip exemplifies an emerging class of high-efficiency optoelectronic devices that utilize electrical power to transform ambient heat into non-equilibrium thermal radiation.[23–27] Our approach enables bootstrap-free, low-input-voltage boost conversion with constraints on efficiency and power density mirroring those of $O(W/mm^2)$ high-power LEDs.

**Conceptual overview**

Fig. 1(a) depicts the basic operating principle of our voltage boost conversion system and highlights key features of our approach. When a forward bias is applied to the LED, it emits photons into a transparent, high-index substrate. These photons undergo diffuse scattering and collectively illuminate multiple photovoltaic cells (PVs) that are optically in contact with the same substrate but at a different point laterally. Because the current through an LED or PV is an exponential function of its voltage in dark conditions, if a PV's photocurrent is a fraction $\eta_{CQE}$ (i.e., the coupling quantum efficiency) of the LED's forward current, the open-circuit voltage of the PV $V_{OC,single-PV}$ far exceeds the same fraction of the original LED drive voltage $V_{LED} \times \eta_{CQE}$. Therefore, a circuit that combines $N$ such PVs in series can supply a boosted voltage $V_{OC,PV-string} = N \times V_{LED} \times \eta_{CQE} > V_{LED}$.

Recent work utilizing this voltage gain mechanism includes a board-scale demonstration and theoretical studies of potential high-efficiency circuits[17] while a new type of packaged commercial opto-isolator[18–22] has emerged providing more modest gain and efficiency for applications where optical isolation is critical. Voltage gain has also been achieved through illumination by semiconductor laser diodes, both in direct pursuit of voltage conversion itself at board-scale[28] and opportunistically in the context of reducing Joule heating losses in high power density receivers for optical power transmission[29,30]. These approaches however are subject to fundamental limitations due to Stokes losses arising from the mismatch of the LED and PV active layer bandgaps and optical losses arising from needing to propagate light through low-index regions. To our knowledge, no attempts have been made to monolithically integrate LEDs and PVs on a single die to achieve voltage boost conversion as we demonstrate here.

The converter architecture depicted in Fig. 1(a) combines many Double Hetero-Junction (DHJ) *p-n* diodes fabricated in parallel from a shared epitaxial layer stack. The DHJ's active layer emits photons in a wavelength band where the substrate is transparent and nearly index-matched to the active layer. For the converter to generate significant output current, the total area through which photons can exit the LED and the total area through which photons can enter the PV string must both be substantial. As a result, the area of the PIC in Fig. 1(b) is divided almost equally between one large-area LED diode and many small-area diodes that operate as PVs.

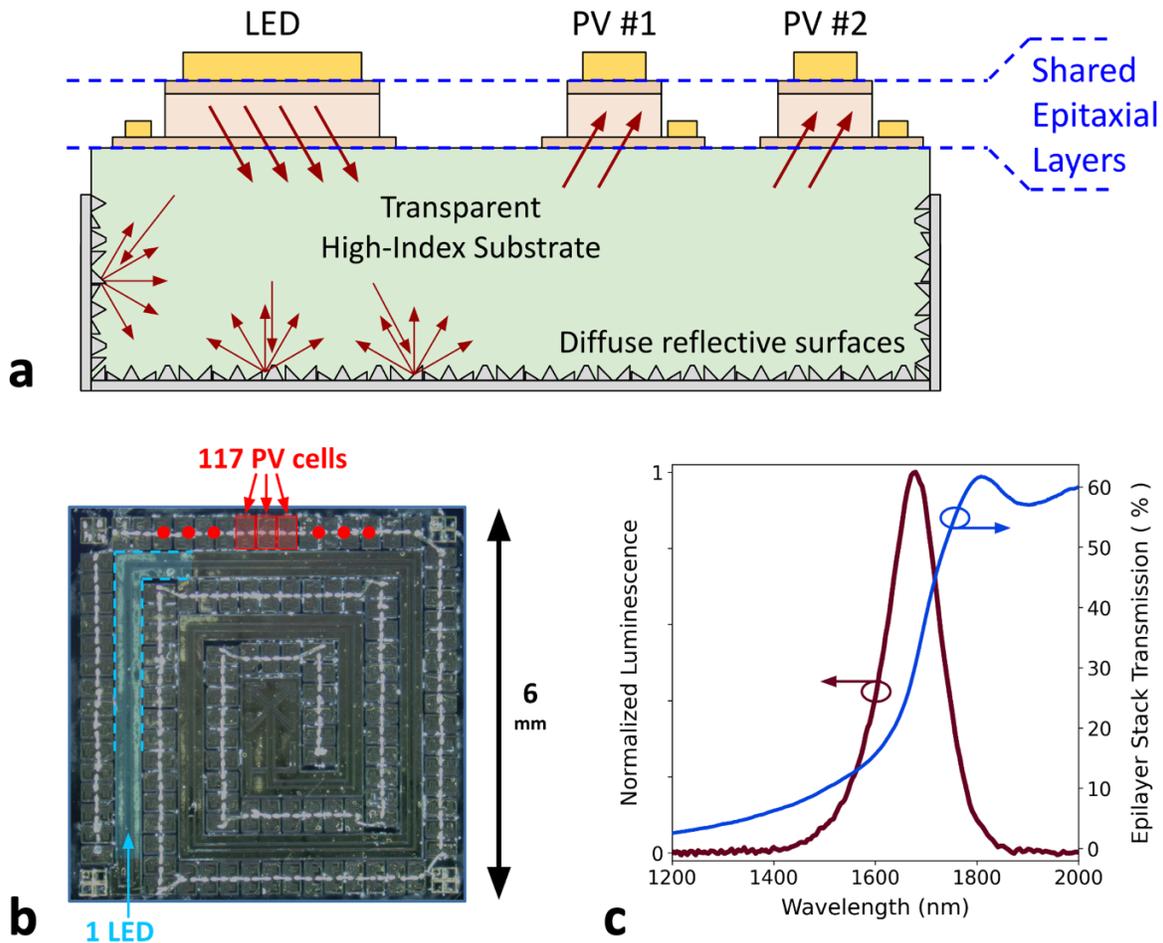

**Fig. 1 | Boost converter architecture and demonstration platform. (a)**, cross-sectional schematic of the voltage boost converter chip. A single larger-area LED (top left of chip) emits photons that illuminate many smaller-area PVs (represented by just two devices in the top right for clarity). The substrate is surrounded by diffuse reflective surfaces that scatter the emitted photons to maximize the fraction converted back to current by a PV. **(b)**, overhead view of the demonstration chip. 117 small PV devices (each 0.1 mm$^2$) surround a single long, spiral LED (9 mm$^2$) and combine to cover ≈60% of the die's top surface within the 36 mm$^2$ circuit reticle (see Supplementary Information §SI-1 for details of the chip layout). **(c)**, luminescence and normal-incidence transmission spectra of the shared epilayer stack. The epilayer stack includes a 2µm-thick InGaAs layer whose inter-band transitions dominate photon absorption across the emission spectrum. Transmission at 1800-2000 nm is limited by index contrast at the semiconductor-air boundaries along the measurement path; photon absorption in this band is negligible to within our experimental uncertainty.

A direct consequence of the shared-epitaxy approach is the partial transparency of the PV to the LED's emitted photons as quantified in Fig. 1(c). For architectures prone to the parasitic absorption or leakage of photons, partial transparency severely restricts the coupling efficiency $\eta_{CQE}$, resulting in low efficiency and low gain. As a result, previous multi-die LED-PV voltage converters achieved boost gain by illuminating Silicon ($E_{gap,PV} \approx 1.1$ eV) with AlGaAs photons ($E_{gap,LED} > 1.4$ eV), a choice that limits the maximum power conversion efficiency via Stokes energy loss, precludes both the simplicity of a fully-integrated monolithic design, and results in severe power density limitations at input voltages below 1 Volt[17–22]. In our monolithic architecture, light is coupled from the single, large LED to each small PV through a high index substrate. This allows modes far from normal incidence, whose path lengths through the absorptive PV epilayers are longer, to contribute to $\eta_{CQE}$ and mitigate partial transparency. Moreover, because photons are confined within a low optical loss medium by a reflective top contact and diffusely reflective substrate surfaces, even photons on the long wavelength tail that require multiple vertical passes through the substrate may contribute.

The strict equivalence of the emitter and absorber bandgaps in a shared epitaxy architecture simplifies fabrication, enables buck-boost bidirectionality, improves power conversion efficiency by removing the fundamental Stokes loss contribution, and minimizes the input voltage required to turn on the LED. A potential advantage over established technologies, our boost converter's minimum input voltage is further reduced by our choice of InGaAs ($E_{gap}$ = 0.74 eV) and explicit design of the epi stack for dilute minority carrier densities[31,32]. Together, these design choices enable input voltages well below the LED's bandgap voltage and less than half that in any of Refs. [17–22,28,29].

**Electron and photon transport in a substrate-coupled architecture**

To characterize the advantages of a single-die, monolithic converter architecture (Fig. 1(a)) in which all emitting and absorbing regions are immersion-coupled by a shared substrate wafer, we model two similar simplified DC-DC boost converters that employ our experimental epilayer stack. We compare one converter in a substrate-coupled optical configuration (Fig. 2(a)) with a converter that contains a thick air gap between the LED side and the PV side (Fig. 2(b)). We first simulate the optical transport using the Transfer Matrix Method (TMM) to calculate the probability that photons emitted from the LED active region will be absorbed in the PV active layer ($P_A$), reflected back to the emitter ($P_R$), or lost to parasitic absorption ($P_L = 1 - P_A - P_R$) at each wavelength and angle. Using PL-weighted, angle-averaged values of $P_A$, $P_R$, and $P_L$, we calculate effective radiative recombination coefficients that incorporate the effects of photon recycling and use them in the electrical simulations. The geometry of the optical TMM calculation for the substrate-coupled case is depicted in Fig. 2(a). The thick InP:Fe substrate (n ≈ 3.15, α ≈ 0.15 cm$^{-1}$ at λ ≈ 1600 nm)[33,34] is sandwiched by the DHJ epilayer stack from our experimental demonstration (see §SI-2) with *n*-side adjacent the substrate on both sides. The stack is composed entirely of InP and InGaAs ($n_{InGaAs}$ ≈ 3.58)[35,36] with the ideal mirror modeled as a semi-infinite layer with ε = -10$^{12}$ + 0i. For details of optical constants used, see §SI-6.

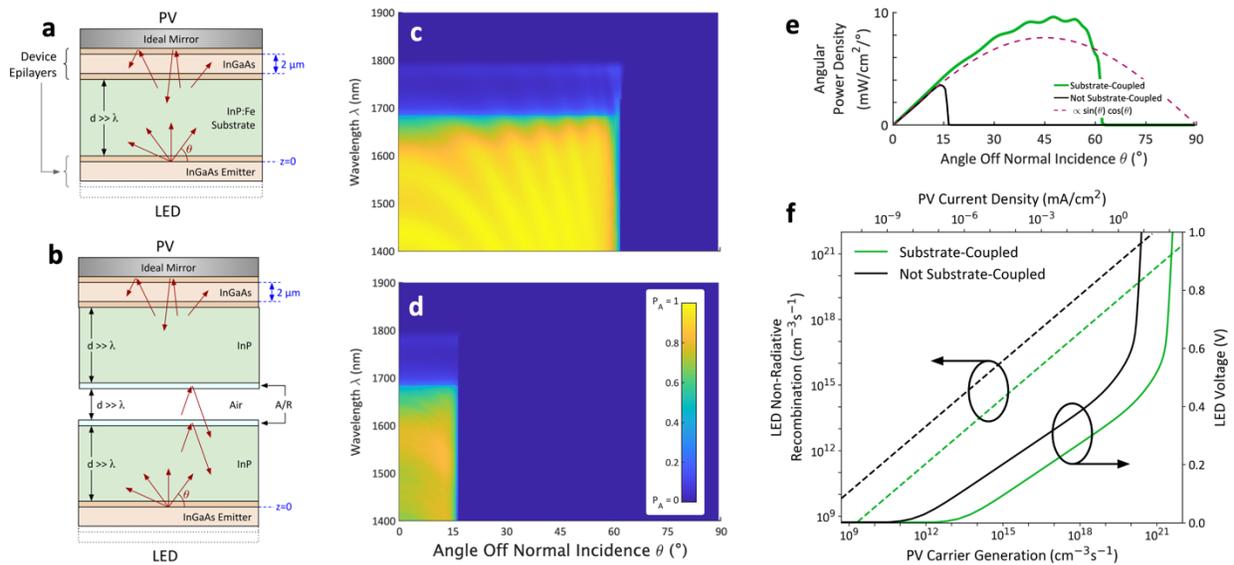

**Figure 2: Impact of the shared-substrate, shared-epilayer platform. (a)-(b)**, diagrams of the 1D multi-layer stacks used in the optical segment of a numerical model used to explore the benefits of coupling the emitting LED to the absorbing PVs directly through a high-index medium such as an InP wafer. In the diagram, the InGaAs layers and the ones immediately surrounding them are modeled with the actual as-grown epilayer stack used in our experimental demonstration; details of the layer stack, particularly those whose purpose relates primarily to electron transport, are suppressed here to simplify the diagram. Greater detail can be found in Fig. 3(f) and §SI-2. **(c)-(e)**, results of a TMM simulation of (**a**) and (**b**) before the inclusion of an exogenous $P_L$, later taken as 5% for the electrical simulations. In the substrate-coupled case (**c**), the angular extent of the photonic phase space available to transfer optical energy is much greater than in the non-substrate-coupled case (**d**), as evident in the angular power density

plot in (**e**). (**f**), the ultimate impact of substrate coupling on electrical characteristics. At a given low input voltage (right y-axis), substrate-coupling provides a 60-fold increase in converter's maximum output current and a 50-fold reduction in non-radiative recombination losses (left y-axis).

The non-substrate-coupled setup is depicted in Fig. 2(b) and is identical to the substrate-coupled case with the addition of an air layer in the middle of the substrate connected to the InP by two 300-nm-thick anti-reflection coatings (ARCs) with $n_{ARC}(\lambda) = \text{sqrt}(n_{InP}(\lambda))$. To arrive at the input parameters for the electron transport model, we take $P_L(\lambda) = 5\%$ for both configurations. The results of the optical TMM calculation are shown in Fig. 2(c), (d), and (e); all quantities in these figures are averaged over both polarizations. In both Fig. 2(c) and (d), we see a large probability of absorption in the PV delineated by two clear edges, one spectral and one angular. The spectral edge descends from the inter-band absorption edge of direct-bandgap InGaAs. The angular dependence descends from the refractive index of the lowest-index layer or, equivalently, the layer that supports the least étendue $\xi \sim n^2$. In the substrate-coupled case, this layer is the InP substrate and $P_A \approx 0$ for $\theta > \theta_{crit,InP} \approx 62º$ because of TIR at the InGaAs-InP interface. Similarly, in the non-substrate-coupled case, the étendue bottleneck is the Air layer and $P_A \approx 0$ for $\theta > \theta_{crit,Air} \approx 16º$ because of TIR at the lower InGaAs-InP, InP-ARC, and ARC-Air interfaces. With careful design of a multi-layer ARC, transmission from the LED epilayers to the PV epilayers can be high on-axis and the short-wavelength $P_A$ can approach 1, but the constraint imposed by étendue is fundamental for layers of thickness $d \gg \lambda$. Consequently, the étendue bottleneck having $n_{InP}^2 \approx 10$ rather than $n_{Air}^2 = 1$ results in 10× as many photon modes contributing to $P_A$ even with an ideal multi-layer ARC.

However, the benefit of substrate coupling as evidenced by the PL-weighted, angle-averaged value of $P_A$ is even greater than 10× because the angles that can contribute only with substrate coupling (i.e. $\theta_{crit,Air} > \theta > \theta_{crit,InP}$) have longer path lengths through the PV-side InGaAs. This fact is shown in Fig. 2(e), in which the power per unit polar angle absorbed by the PV when 1 W/cm² of luminescence is emitted upward by the LED. The substrate-coupled case exceeds the $\sim \sin(\theta)\cos(\theta)$ extrapolation of the non-substrate-coupled power to $\theta > \theta_{crit,Air}$. We find the PL-weighted, angle-averaged probability of a photon generated in the LED active layer being absorbed in the PV active layer $P_A$ to be ≈0.37 and ≈0.03 for the substrate-coupled and non-substrate-coupled cases, respectively.

We combine these simulations in Fig. 2(f) and highlight the substantial reduction in both non-radiative recombination and increase in PV carrier generation at a given voltage that is enabled by our substrate-coupled approach. In a boost converter, the current produced at the converter output would be split among PVs that operate electrically in series. Thus, our result is agnostic to design choices like the layout and number of PVs in the converter. As an example, if a 1 cm$^2$, 100-PV boost converter were producing output power with a short-circuit current of 1 mA, then the PV current density would be 100 mA/cm$^2$. Our simulations indicate that this 1 mA condition could be achieved with an input voltage of ≈0.3 Volts (less than 50% InGaAs photon voltage ≈0.74 Volts) with a substrate-coupled architecture while the non-substrate-coupled converter would be limited to < 16 µA. The factor of 64× separating those current levels is the combined impact of a larger effective bimolecular recombination rate leading to a larger number of photons being generated (a factor of ≈5×) and a larger portion of those photons being able to reach the PV due to substrate-coupling (a factor of ≈12×). Moreover, achieving similar performance is not merely a matter of increasing the input voltage at the cost of moderately diminished gain. If 1 mA were required, the additional ≈110 mV required by the non-substrate-coupled converter would require the LED to maintain an ≈70× times larger concentration of excited minority electrons, leading to ≈50 times more non-radiative recombination in our model.

**Experimental demonstration of boost gain by a monolithic, chip-scale, DC-DC voltage converter**

To experimentally validate our voltage boost converter architecture, we designed, fabricated, and tested a single-die monolithically integrated optoelectronic chip. The chip is shown in Fig. 3 and includes several PV strings of different lengths that can be simultaneously illuminated by a single, large-area LED. Although power is simultaneously generated in the strings, each string is characterized as an isolated converter circuit driven by the same input with all other devices on the chip kept at open-circuit and treated as parasitic absorbers. Each 0.1 mm$^2$ PV cell occupies 500 × 280 µm cell within the full 6 × 6 mm converter circuit (Fig. 3(a)) in a two-level electrically isolated mesa. The taller mesa area contains the InGaAs active region while the lower mesa level area is etched back to an *n*-type etch stop epilayer buried under the active region. The metal pad atop the taller mesa is the PV's *p*-contact and is wire bonded to an *n*-contact pad that resides on the lower mesa of an adjacent PV. Ti/Pt/Au metal is used for both the *n*- and *p*-type contacts[37,38]. The top (+) contact covers as much of the active mesa area as possible given fabrication constraints; this increases the efficiency of photon collection from the substrate and photon extraction to the substrate in the PV and LED devices, respectively (Fig. 3(b)). The transparent high-index InP:Fe substrate, has an absorption length of ≈200× the wafer thickness and whose interface with InGaAs leads photons escaping the active layer to face a manageable contrast of $(n_{InGaAs} - n_{InP}) / n_{InGaAs} \approx 12\%$.

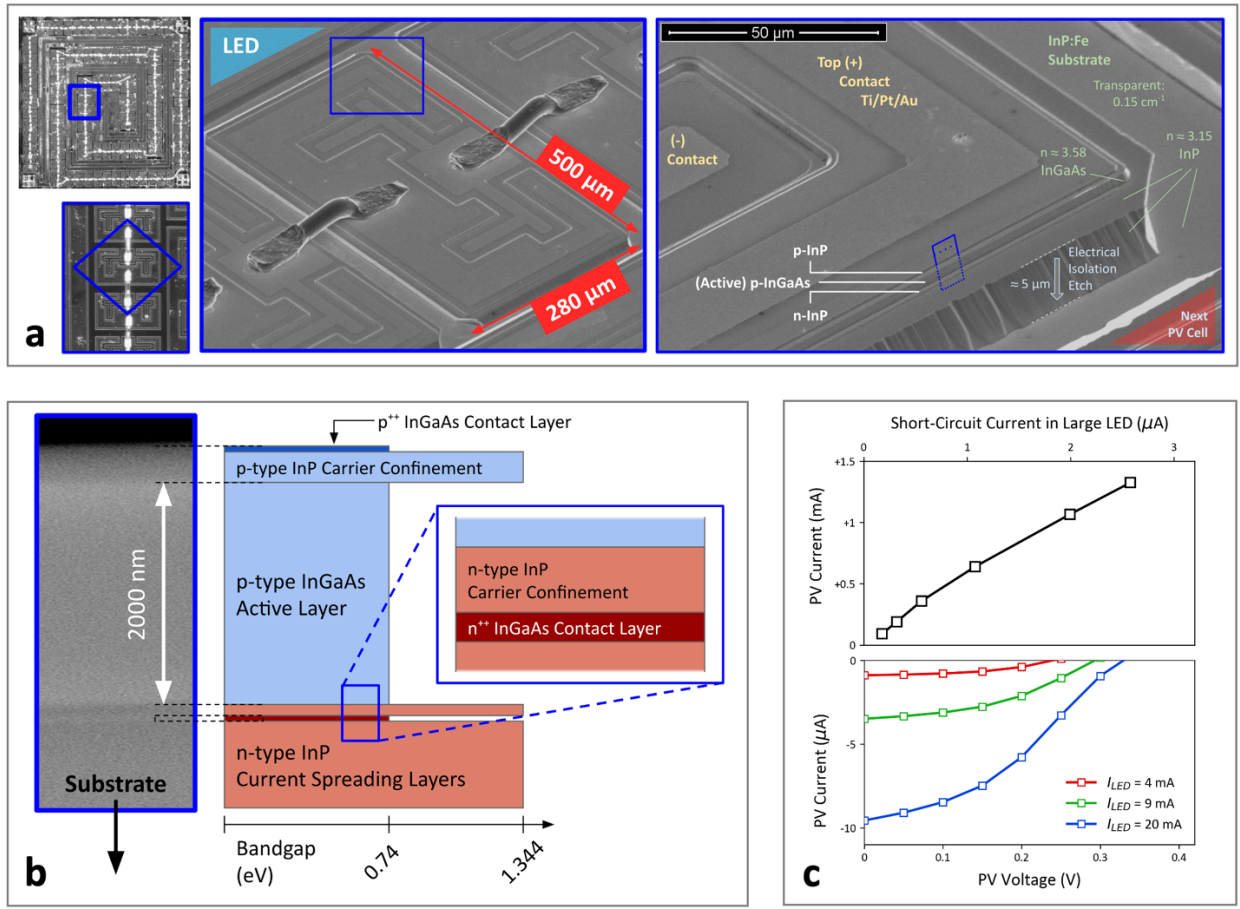

**Figure 3: Fabricated devices demonstrate the basic physical mechanisms required for voltage boost conversion. a**, progressively higher magnification images of the fabricated chip ranging from a 6×6-mm field of view showing the entire circuit down to a ≈100 μm-wide field of view showing the corner of one PV cell. The leftmost pair are optical microscope images while the two larger images are low-magnification SEM images. **b**, a high-magnification SEM of the cross-section of the as-grown epilayer stack used to fabricate our chip and an aligned diagram of the epilayer stack depicting the doping and bandgap of various important layers. Red (blue) denotes *n*-type (*p*-type) material while the darkness

of tone indicates doping magnitude; further details on the epilayer stack can be found in §SI-2. **c**, four-terminal electrical data from a pair of diodes, a large LED and a small PV cell like the ones in **a**, monolithically fabricated side-by-side from shared InGaAsP epitaxy. As expected, driving the large LED generates a photovoltaic response in the small PV (bottom), while driving a forward current through the small PV (i.e., operating it as an LED) generates a short-circuit current $I_{sc}$ in the large LED (top).

When the larger LED is used as a detector, driving the PV cell into forward bias causes photons to be emitted into the substrate (Fig. 3(c), top). Though we do not explore it further here, this behavior supports the potential for buck conversion in this architecture. When illuminated by the large-area LED, the device exhibits a photovoltaic response (Fig. 3(c), bottom). These measurements show that our devices exhibit the necessary opto-electronic processes to demonstrate voltage boost conversion.

Driving the LED to illuminate the PV cells, we measure a peak gain of 3.8× from a string of 24 PVs and a peak output power of 4.9 µW from a non-overlapping string of 7 PVs (Fig. 4). Fig. 4(a) identifies three such series-connected strings of PVs whose electrical responses to an input voltage $V_{LED}$ is photovoltaic power generation at an elevated voltage $V_{PV\text{-}string} > V_{LED}$, allowing each to be analyzed as a distinct voltage boost converter. We next measured the full current-voltage characteristic of the 24-diode PV string under a range of drive conditions for the large, spiral LED (Fig. 4(b)). We measure the voltage and boost gain for both 29- and 24-PV strings in the chip (Fig. 4(c)) with peak gain values of 3.7 and 3.8 observed respectively. We next calculated the performance characteristics of the 24- and 7-diode strings (Fig. 4(d)). We find peak powers of 4.4 and 4.9 µW respectively. Since output and input power scale similarly our overall power conversion efficiency is stable across a broad range of power levels.

Overall device performance could be improved through a range of strategies. First, the limited photovoltaic fill factor observed at higher input powers is likely caused by light-induced electrical shunts. Data shown Fig. 3(c) indicates single-PV shunt resistances of 1.5 MΩ and 120 kΩ are present at $I_{LED} = 0$ and 20 mA, respectively, providing evidence of light-induced conductive paths in the chip. Furthermore, ambient light in the lab was found to reduce PV strings' open-circuit voltages. Together these observations suggest light-tight packaging could improve converter performance. Series resistances at the probe-contact interface represent another area for improvement. Difficulty keeping the sample surface clean in the presence of delicate wire bonds led to parasitic series resistances O(10 Ω), suggesting ≈ 20% reductions of input voltage and matching increases in gain should be possible through characterization improvements alone.

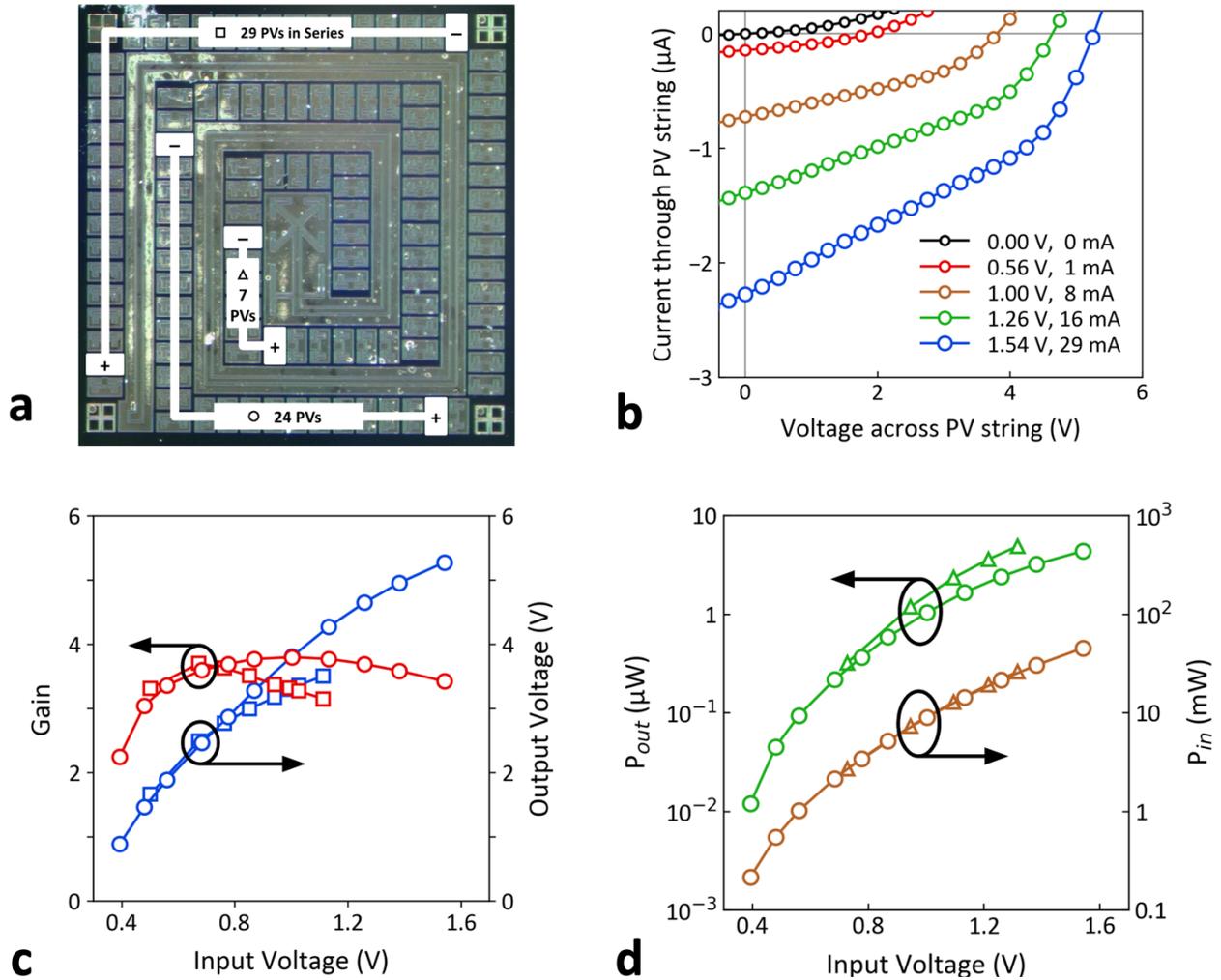

**Fig. 4 | Demonstration chip performance. a**, image of demonstration chip (shown prior to wire bonding for clarity) indicating the locations of 3 strings of PVs whose responses to illumination by a monolithically integrated LED are shown in **b-d**. Data from the 29-, 24-, and 7-diode strings, wire bonded in series and probed at the ends, are indicated by square, circle, and triangle markers, respectively. **b**, current-voltage characteristics of the 24-diode PV string while the LED is driven at between 0 and 1.54 Volts. **c**, performance characteristics at twe open-circuit point (where output voltage and gain are maximized) for the two longer PV strings labeled in **a**. Peak gain values of 3.7 and 3.8 are observed for the 29-diode and 24-diode strings, respectively. **d**, performance characteristics at the max-power point. Peak powers of 4.4 and 4.9 µW were observed for the 24-diode and 7-diode strings, respectively; larger

values are possible, but excitations were limited to < 50 mW to avoid thermal runaway and associated damage to the LED. Output power and input power scale similarly, indicating a relatively stable conversion efficiency across a range of power levels.

The sub-optimal thickness of the p-contact metal pads limited the LED drive conditions to < 50 mW, indicating another potential opportunity to increase output voltage and gain. Based on models that we expand on in the SI, the reasons for the gain peaking around 1 Volt appears to relate to the LED. The gain falls below 0.6 Volts because of SRH recombination, which reduces the LED's internal quantum efficiency; in this parasitic process, the LED's large, unpassivated sidewall area is likely a contributing factor. The reason for gain falling at high voltage is the LED's series resistance, which we trace back to a combination of characterization limitations, current spreading on the *p*-side, and vertical transport across hetero-barriers in the epilayer stack. These observations suggest that implementing simple parametric device optimization and common fabrication steps could extend the range of efficient voltage boost conversion up toward power densities more typical of power electronic ICs and down toward low input voltages where bootstrap-free startup is a challenge for existing chip-scale boost converters.

**Conclusions**

The voltage boost gain from a monolithic PIC shown here presents the possibility of a new generation of power electronic chips capable of harnessing energy resources and operating in electromagnetic environments inaccessible to incumbent technologies. Low input voltage measurements on a fully integrated, bootstrap-free, voltage boost converter demonstration chip indicate that its InP-based platform could leverage telecom PIC fabrication infrastructure to manufacture components that can, either as a charge pump bootstrap or a standalone system, power digital and low-noise analog electronics using intermittent energy resources harvested at very low voltages. Key engineering challenges will include minimizing the parasitic absorption and leakage of photons to ambient, developing a robust, monolithic interconnect layer, achieving uniform illumination of the PVs despite current spreading resistance in the LED, and improving the LED's quantum efficiency at low voltage by optimizing the epilayer stack and fabrication. Experimental evidence that LEDs can produce light at deep-sub-thermal voltages ($V \ll k_BT/q$) has prompted design and demonstration work pointing to a striking symmetry between LEDs and PVs optimized for efficiency in the low voltage limit, indicating a potential for record-breaking low input voltage converters based on our monolithic, shared-epilayer demonstration platform[27,31,32,39–42].

      Decarbonizing global energy infrastructure has brought new urgency to the field of power electronics and new opportunities for large-scale voltage conversion technologies to interconnect new forms of supply like solar and wind with new forms of demand like electric vehicles and datacenters. Meanwhile, the incentive to harvest energy in-situ to drive microelectronic systems has created similar opportunities at very small scales. The past 30 years has seen at least two semiconductor optoelectronic energy technologies displace heavily entrenched but limited incumbent technologies: solar photovoltaics and LED lighting are rapidly expanding as the limitations and negative externalities of hydrocarbon combustion and incandescent lighting become more undeniable. Our results point to voltage conversion as a potential candidate to join the ranks of baseload power generation and general-purpose lighting as the next energy technology to transition to a semiconductor optoelectronic solution by lifting pervasive limitations on voltage scale, electromagnetic compatibility, and miniaturization.

**Acknowledgements**

This material is based upon work supported by the Defense Advanced Research Projects Agency (Award #W911NF2110345).

# References

1. Tarzamni, H., Gohari, H. S., Sabahi, M. & Kyyrä, J. Nonisolated High Step-Up DC–DC Converters: Comparative Review and Metrics Applicability. *IEEE Trans. Power Electron.* **39**, 582–625 (2024).

2. Atkar, D. D., Chaturvedi, P., Suryawanshi, H. M., Nachankar, P. P. & Yadeo, D. Optimal Design of Solid State Transformer-Based Interlink Converter for Hybrid AC/DC Micro-Grid Applications. *IEEE J. Emerg. Sel. Top. Power Electron.* **10**, 3685–3696 (2022).

3. Dickson, J. F. On-chip high-voltage generation in MNOS integrated circuits using an improved voltage multiplier technique. *IEEE J. Solid-State Circuits* **11**, 374–378 (1976).

4. Ballo, A., Grasso, A. D. & Palumbo, G. A Current-Efficient Pseudo-3D Regulated Dickson Charge Pump. *IEEE Trans. Power Electron.* **39**, 9878–9890 (2024).

5. Zare, F. *Power Electronics Education, Electronic-Book*. (PEEEB, Brisbane, Australia, 2010).

6. Erickson, R. W. & Maksimović, D. *Fundamentals of Power Electronics*. (Springer International Publishing, Cham, 2020). doi:10.1007/978-3-030-43881-4.

7. Sullivan, C. R., Reese, B. A., Stein, A. L. F. & Kyaw, P. A. On size and magnetics: Why small efficient power inductors are rare. in *2016 International Symposium on 3D Power Electronics Integration and Manufacturing (3D-PEIM)* 1–23 (IEEE, Raleigh, NC, USA, 2016). doi:10.1109/3DPEIM.2016.7570571.

8. Rongxiang Wu, Sin, J. K. O. & Hui, S. Y. A novel silicon-embedded coreless transformer for isolated DC-DC converter application. in *2011 IEEE 23rd International Symposium on Power Semiconductor Devices and ICs* 352–355 (IEEE, San Diego, CA, 2011). doi:10.1109/ISPSD.2011.5890863.

9. Fuketa, H. *et al.* An 85-mV input, 50-µs startup fully integrated voltage multiplier with passive clock boost using on-chip transformers for energy harvesting. in *ESSCIRC 2014 - 40th European Solid State Circuits Conference (ESSCIRC)* 263–266 (IEEE, Venice Lido, Italy, 2014). doi:10.1109/ESSCIRC.2014.6942072.

10. Liang, W., Raymond, L. & Rivas, J. 3-D-Printed Air-Core Inductors for High-Frequency Power Converters. *IEEE Trans. Power Electron.* **31**, 52–64 (2016).

11. Flowers, P. F., Reyes, C., Ye, S., Kim, M. J. & Wiley, B. J. 3D printing electronic components and circuits with conductive thermoplastic filament. *Addit. Manuf.* **18**, 156–163 (2017).

12. Carlson, E. J., Strunz, K. & Otis, B. P. A 20 mV Input Boost Converter With Efficient Digital Control for Thermoelectric Energy Harvesting. *IEEE J. Solid-State Circuits* **45**, 741–750 (2010).

13. Ramadass, Y. K. & Chandrakasan, A. P. A Battery-Less Thermoelectric Energy Harvesting Interface Circuit With 35 mV Startup Voltage. *IEEE J. Solid-State Circuits* **46**, 333–341 (2011).

14. Xiwen Zhang & Hoi Lee. Gain-Enhanced Monolithic Charge Pump With Simultaneous Dynamic Gate and Substrate Control. *IEEE Trans. Very Large Scale Integr. VLSI Syst.* **21**, 593–596 (2013).

15. Damaschke, J. M. Design of a Low-Input-Voltage Converter for Thermoelectric Generator. *IEEE Trans. Ind. Appl.* **33**, (1997).

16. Kimball, J. W., Flowers, T. L. & Chapman, P. L. Issues with low-input-voltage boost converter design. in *2004 IEEE 35th Annual Power Electronics Specialists Conference (IEEE Cat. No.04CH37551)* 2152–2156 (IEEE, Aachen, Germany, 2004). doi:10.1109/PESC.2004.1355452.

17. Zhao, B., Assawaworrarit, S., Santhanam, P., Orenstein, M. & Fan, S. High-performance photonic transformers for DC voltage conversion. *Nat. Commun.* **12**, 4684 (2021).

18. Infineon Technologies AG. Infineon PVI5080NPbF Datasheet. (2017).

19. MH GoPower Company Limited. MH GoPower YMH-HH2A42 Datasheet - Isolated Gate Driver. (2018).

20. Panasonic Corporation. Panasonic APV1111GVY and APV3111GVY Datasheet - PV MOSFET driver high power type. (2021).

21. Vishay Semiconductors. Vishay VO1263AAC Datasheet - Dual PV MOSFET driver solid-state relay. (2023).

22. Toshiba Electronic Devices & Storage Corporation. Toshiba TLX9910 Datasheet - IR LED and Photo Diode Photocoupler. (2023).

23. Tauc, J. The share of thermal energy taken from the surroundings in the electro-luminescent energy radiated from ap-n junction. *Czechoslov. J. Phys.* **7**, 275–276 (1957).

24. Santhanam, P., Gray, D. J. & Ram, R. J. Thermoelectrically Pumped Light-Emitting Diodes Operating above Unity Efficiency. *Phys. Rev. Lett.* **108**, 097403 (2012).

25. Hurni, C. A. *et al.* Bulk GaN flip-chip violet light-emitting diodes with optimized efficiency for high-power operation. *Appl. Phys. Lett.* **106**, 031101 (2015).

26. Xiao, T. P., Chen, K., Santhanam, P., Fan, S. & Yablonovitch, E. Electroluminescent refrigeration by ultra-efficient GaAs light-emitting diodes. *J. Appl. Phys.* **123**, 173104 (2018).

27. Sadi, T., Radevici, I. & Oksanen, J. Thermophotonic cooling with light-emitting diodes. *Nat. Photonics* **14**, 205–214 (2020).

28. Wilkins, M. M. *et al.* Ripple-Free Boost-Mode Power Supply Using Photonic Power Conversion. *IEEE Trans. Power Electron.* **34**, 1054–1064 (2019).

29. Fafard, S. *et al.* Ultrahigh efficiencies in vertical epitaxial heterostructure architectures. *Appl. Phys. Lett.* **108**, 071101 (2016).

30. Putra, E. P. *et al.* Technology update on patent and development trend of power over fiber: a critical review and future prospects. *J. Photonics Energy* **13**, (2023).

31. Heikkilä, O., Oksanen, J. & Tulkki, J. Ultimate limit and temperature dependency of light-emitting diode efficiency. *J. Appl. Phys.* **105**, 093119 (2009).

32. Santhanam, P. *et al.* Controlling the dopant profile for SRH suppression at low current densities in λ ≈ 1330 nm GaInAsP light-emitting diodes. *Appl. Phys. Lett.* **116**, 203503 (2020).

33. Pettit, G. D. & Turner, W. J. Refractive Index of InP. *J. Appl. Phys.* **36**, 2081–2081 (1965).

34. Adachi, S. *Physical Properties of III-V Semiconductor Compounds: InP, InAs, GaAs, GaP, InGaAs, and InGaAsP*. (Wiley, New York, 1992).

35. Dinges, H. W., Burkhard, H., Lösch, R., Nickel, H. & Schlapp, W. Refractive indices of InAlAs and InGaAs/InP from 250 to 1900 nm determined by spectroscopic ellipsometry. *Appl. Surf. Sci.* **54**, 477–481 (1992).

36. Kowalsky, W., Wehmann, H.-H., Fiedler, F. & Schlachetzki, A. Optical Absorption and Refractive Index near the Bandgap for InGaAsP. *Phys. Status Solidi A* **77**, K75–K80 (1983).

37. Lin, J. C., Yu, S. Y. & Mohney, S. E. Characterization of low-resistance ohmic contacts to *n* - and *p* -type InGaAs. *J. Appl. Phys.* **114**, 044504 (2013).

38. Yu, J. S., Kim, S. H. & Kim, T. I. PtTiPtAu and PdTiPtAu ohmic contacts to p-InGaAs. in *Compound Semiconductors 1997. Proceedings of the IEEE Twenty-Fourth International Symposium on Compound Semiconductors* 175–178 (IEEE, San Diego, CA, USA, 1997). doi:10.1109/ISCS.1998.711608.

39. Lehovec, K., Accardo, C. A. & Jamgochian, E. Light Emission Produced by Current Injected into a Green Silicon-Carbide Crystal. *Phys. Rev.* **89**, 20–25 (1953).

40. Ban, D. *et al.* Optimized GaAs∕AlGaAs light-emitting diodes and high efficiency wafer-fused optical up-conversion devices. *J. Appl. Phys.* **96**, 5243–5248 (2004).

41. Santhanam, P., Huang, D., Ram, R. J., Remennyi, M. A. & Matveev, B. A. Room temperature thermo-electric pumping in mid-infrared light-emitting diodes. *Appl. Phys. Lett.* **103**, 183513 (2013).

42. Tripathi, T. S., Radevici, I., Dagyte, V., Sadi, T. & Oksanen, J. Improving the Efficiency of GaInP/GaAs Light Emitters Using Surface Passivation. *IEEE Trans. Electron Devices* **67**, 3667–3672 (2020).